\documentclass{pnastwo}
\usepackage{color}

\usepackage{graphicx}
\usepackage{cite}

\usepackage{amssymb,amsfonts,amsmath}

\newcommand{\Tg}{T_{\mathrm{g}}}
\newcommand{\Tag}{T_{\mathrm{ag}}}

\newcommand{\po}{p_{\mathrm{o}}}
\newcommand{\df}{d_\mathrm{f}}
\newcommand{\tbeta}{\tilde{\beta}}

\newcommand{\To}{T_{\mathrm{o}}}

\newcommand{\tobs}{t_{\mathrm{obs}}}

\url{}

\footlineauthor{Limmer, et al.}
\volume{}
\issuenumber{}

\begin{document}

\title{Theory of amorphous ices}

\author{David T. Limmer\affil{1}{Department of Chemistry, University of California, Berkeley, 94609} and
David Chandler\affil{1}{University of California, Berkeley} }

\contributor{Submitted to Proceedings of the National Academy of Sciences of the United States of America}

\maketitle

\begin{article}
\begin{abstract}
We derive a phase diagram for amorphous solids and liquid supercooled water and explain why the amorphous solids of water exist in several different forms.  Application of large-deviation theory allows us to prepare such phases in computer simulations.  Along with nonequilibrium transitions between the ergodic liquid and two distinct amorphous solids, we establish coexistence between these two amorphous solids.  The phase diagram we predict includes a nonequilibrium triple point where two amorphous phases and the liquid coexist.   While the amorphous solids are long-lived and slowly-aging glasses, their melting can lead quickly to the formation of crystalline ice.  Further, melting of the higher density amorphous solid at low pressures takes place in steps, transitioning to the lower density glass before accessing a nonequilibrium liquid from which ice coarsens. 
\end{abstract}


\dropcap{A}morphous ices are nonequilibrium, low temperature phases of water~\cite{loerting2006amorphous, angell2004amorphous, mishima1985apparently}. These phases lack long range order and their properties are fundamentally dependent on the protocols by which they are prepared~\cite{Debenedetti:2003p813, angell1995formation}.  They are molecular glasses that exhibit a variety of reproducible behaviors, including transitions between different amorphous states.  This paper provides quantitative analysis and numerical simulation of this polyamorphism and predicts a nonequilibrium phase diagram, offering explanations of previous experimental observations~\cite{mishima1985apparently,loerting2006amorphous,elsaesser2010reversibility,Mishima:2000p4162, chen2011high,amann2013water} and possibly guiding future experiments on supercooled water. 

Our treatment can be applied generally in many cases where there is interest in comparing phase behaviors of non-equilibrium glasses with those of equilibrium liquids and crystals.  For water in particular, however, our results bear on whether observed nonequilibrium polyamorphism can be interpreted as evidence of more than one distinct liquid phase of water.  It is a topic of current interest and controversy.  There are no direct measurements of two-liquid behavior in water, but the low-temperature critical point that would accompany such behavior has been offered as an explanation for unusual properties of liquid water, such as maxima in various response functions~\cite{Debenedetti:2003p813, poole1994effect}, and molecular simulation results are often cited as supporting this theoretical idea, e.g., Refs.\cite{li2013liquid,poole2013free,liu2012liquid,kesselring2012nanoscale}.  But water anomalies can be explained with models for which there is demonstrably only one liquid phase\cite{holten2013nature}, and seemingly rigorous equilibrium analysis of various simulation models argues against cold water exhibiting the existence of two distinct liquids\cite{Limmer:2011p134503,limmer2013putative}.  Rather, it seems that an illusion of two-liquid behavior in simulation coincides with coarsening of ice, and this paper shows how arresting those fluctuations yields a multitude of nonequilibrium amorphous solids.

\section{Phenomenology}
\begin{figure}[t]
\begin{center}
\includegraphics[width=8.0 cm]{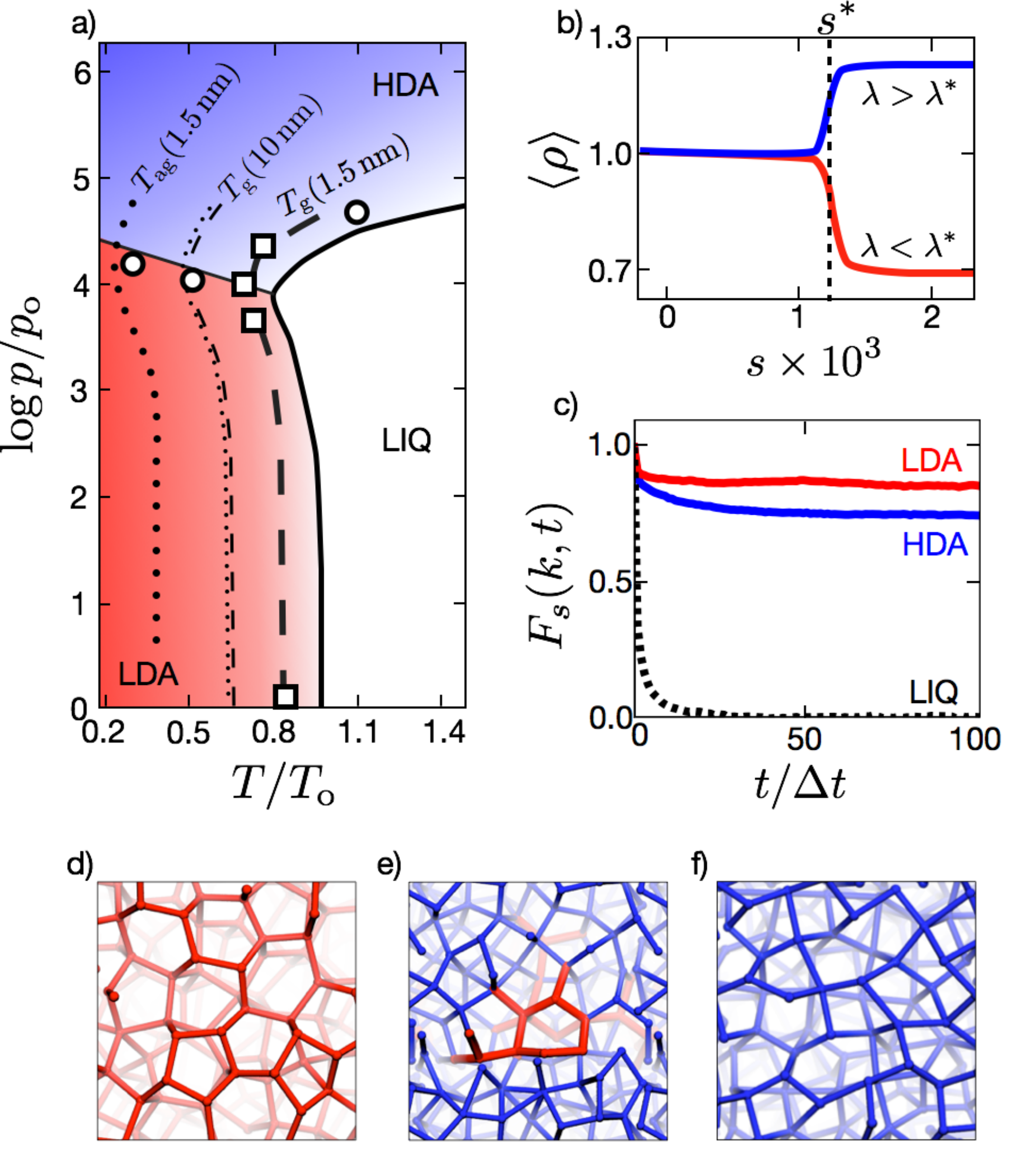}
\caption{\label{Fi:noneqPD}Liquid, LIQ, and nonequilibrium high- and low-density amorphous solids, HDA and LDA.  (a) Pressure-temperature phase diagram for water, with the liquid onset temperature line (solid), glass transition lines (dashed) and apparent glass transition lines (dotted). Squares locate points where nonequilibrium $s$-ensemble calculations locate coexistence between nonequilibrium phases in the mW model.  Circles locate transitions observed in experiments~\cite{loerting2006amorphous,Mishima:2000p4162, chen2011high} and in nonequilibrium relaxation simulations of the mW model. (b) The mean reduced density, $\langle \rho \rangle$, as a function of nonequilibrium control parameters computed for the mW model with the $s$-ensemble near the nonequilibrium triple point. 
(c) Van Hove self correlation functions for the three phases at the liquid's principal wave vector, $k$, all computed for the mW model at conditions near the nonequilibrium triple point.  (d)-(f) Snap shots from simulations, where a bond connecting molecular centers $i$ and $j$ is colored according to the value of $\eta_{ijk}$ averaged over a second neighbor $k$ (see text).  The bond is red if this value is less than 0.1; otherwise, it is blue. A typical configuration of LDA ice is pictured in (d), that of a domain of LDA ice in coexistence with HDA ice is in (e), and that of HDA is in (f).}
\end{center} 
\end{figure}

A phase diagram is drawn in Fig. \ref{Fi:noneqPD}a.  It is partitioned with the onset temperature, $T_\mathrm{o}(p)$, which is the crossover temperature below which liquid-phase dynamics is spatially heterogeneous.  This temperature is an equilibrium material property.  The pressure dependence of $T_\mathrm{o}(p)$ for water has been determined from experimental transport data and computation\cite{limmer2012phase}.  The low-pressure limit of the onset temperature, $T_\mathrm{o}$, coincides with the temperature of maximum density\cite{limmer2013corresponding}.  In the phase diagram, we express temperature, $T$, in units of $\To$.  Similarly, we express pressure, $p$, in units of $\po = -10^{-4}\Delta h/\Delta v$, where $\Delta h$ and $\Delta v$ are, respectively, the molar enthalpy and volume changes upon melting ice at low pressures.  With reduced pressure and temperature units, the phase diagram is reasonably independent of choice of molecular model\cite{limmer2013corresponding}.  Requirements for a suitable model are two fold: 1) The liquid phase exhibits preference for local tetrahedral order, and 2) the liquid freezes into an ice-like crystal with global tetrahedral order.  Values of $\To$ and $\po$, specific lattice structures, absolute melting temperatures and so forth are sensitive to specific choices of molecular model, but all have similar liquid-phase dynamics at temperatures below the onset, and all have ice-melting temperatures reasonably close to the onset\cite{vega2005relation,limmer2013corresponding}.  For experimental water, $\To = 277$\,K and $\po = 0.3$\,bar. 

Occurring as it does below the onset temperature, the dynamics of forming ice at supercooled conditions are complex.  For example, in the initial stages of coarsening at low enough temperatures relatively large density fluctuations occur associated with dynamic heterogeneity.  These fluctuations take place over a range of time scales extending  to milliseconds\cite{Moore:2010p1923,limmer2013putative}, and when viewed on shorter time scales, they are easily confused with the existence of two distinct liquids.  These fluctuations can be arrested and crystallization can be avoided through rapid enough cooling or confinement, producing nonequilibrium amorphous solids of various types with different glass transitions.  For instance, when hyperquenching at a cooling rate, $\nu$, freezing into glass can occur at a temperature $\Tg$, where $1/\nu = |d\tau/dT|_{T=\Tg}$.  Here, $\tau$ stands for the structural relaxation time of the liquid prior to freezing.  Because the rate of increase of $\tau$ increases with decreasing $T$, the glass transition temperature, $\Tg$, decreases with decreasing $\nu$.  Of course, a low enough cooling rate leads to crystallization, not glass formation.  

Importantly, a different $\nu$ and therefore a different $\Tg$ can imply a different type of glass.  This is because the transition at $\Tg$ produces a material with a frozen nonequilibrium correlation length, $\ell_\mathrm{ne}$\cite{keys2013calorimetric}.  This length is the mean-free path between excitations at the glass transition.  (``Excitations'' are defined precisely in the next section.)  Aging or structural relaxation occurs through coupling excitations, the closer the excitations the more frequent the coupling.  In the liquid, $T>\Tg$, the distribution of lengths between excitations is exponential, like that of an uncorrelated gas of particles.  Dynamics in that case takes place hierarchically, with the fastest and slowest time-scales dictated by the domains with smallest and largest $\ell$, respectively.  By contrast, in the glass, $T<\Tg$, the distribution of $\ell$ is non-exponential with a dominant and most-probable length, $\ell_\mathrm{ne}$, and there is a single activation energy associated with that dominant length. As $\Tg$ decreases with decreasing cooling rate, $\ell_\mathrm{ne}$ grows, and a larger length implies a greater stability of the glass.  In particular, the glass formed with a specific $\ell_\mathrm{ne}$ can be cooled far below its $\Tg$, and when it is then heated slowly, it looses its stability at an apparent glass transition temperature $\Tag$, where $\Tag < \Tg$.  The difference $\Tg - \Tag$ grows as $\ell_\mathrm{ne}$ decreases (or equivalently, as $\nu$ increases)\cite{limmer2013communication}.  

The distinction between $\Tg$ and $\Tag$ is important for water precisely because rapid cooling is required to avoid crystallization of that material.  We illustrate the behavior for water in Fig. \ref{Fi:noneqPD}a.  The cooling rate required to produce $\ell_\mathrm{ne} \approx 1.5$\,nm would be $10^8$\,K/s, and that required to produce $\ell_\mathrm{ne} \approx 5$\,nm would be $10^{4}$\,K/s.  The former is slightly faster than usually estimated for typical experimental hyperquenching rates, while the latter is somewhat slower than what would be needed to avoid crystallization.  To reach $\ell_\mathrm{ne} \approx 10$\,nm would require an even slower $\nu \approx 0.1$\,K/s.  A procedure other than straightforward cooling would be needed to produce amorphous solids of water with $\ell_\mathrm{ne} \approx 5$ or 10\,nm.  

The dependence of $\Tg$ and $\Tag$ upon $\nu$ emphasizes that these temperatures are nonequilibrium properties, and their projections onto a $p$-$T$ plane depend implicitly upon the protocol by which the system is driven from equilibrium. The distinction between $\Tg$ and $\Tag$ has been noted by Angell\cite{yue2004clarifying}, but without the connection to $\ell_\mathrm{ne}$.  Formulas with this connection predict the dashed and dotted lines in Fig. \ref{Fi:noneqPD}a.  They are derived elsewhere~\cite{keys2013calorimetric, limmer2013communication} and summarized in our Supporting Information (\emph{SI}).  The formulas allow us to interpret phenomena that have been observed experimentally, and they allow us to anticipate phenomena examined later in this paper.  
 
To begin, notice that $\Tg$ and $\Tag$ are non-monotonic functions of pressure.  This variation reflects the non-monotonic variation of the onset temperature, which in turn reflects a well-known fact about water:  at low pressures, transformations of liquid water to more ordered states (whether to ice or to supercooled liquid) occur with decreases in density, while at high pressures, they occur with increases in density.   A line passing through the locus of minima in $\Tg$ partitions the high- and low-pressure regimes in Fig. \ref{Fi:noneqPD}a.  As the local structure of the HDA region is necessarily distinct from that of the LDA region, there is a possibility of a nonequilibrium transition between the two. This transition occurs in the vicinity of the line separating the LDA and HDA regions in Fig. \ref{Fi:noneqPD}a.  Indeed, as noted by the circles in that figure, a HDA-LDA transition is observed experimentally close to that line.  The transition cannot persist into the liquid because fluctuations in the liquid remove long-lived distinction between the two\cite{limmer2013putative, elmatad2010finite}.  Determining the nature of the transition and its end point requires further analysis, which we will get to soon. 

Also notice in Fig.~\ref{Fi:noneqPD}a that HDA glass with rather small $\ell_\mathrm{ne}$ has been produced experimentally.  Through cycles of changing $T$ and $p$, or by other means~\cite{elsaesser2010reversibility, amann2013water,nelmes2006annealed}, the stability of that material can be enhanced, possibly producing a material with $\ell_\mathrm{ne} \approx 1.5$ or 2\,nm.  In that case $\Tag \approx 130$\,K.  Such a material could be cooled to a very low temperature and de-pressurized, but still with the high-density structure and nonequilibrium length locked in.  From Fig.~\ref{Fi:noneqPD}a we see that subsequent warming would then cause a transition at a temperature close to $\Tag$, at which point, given its pressure and temperature, the destabilized HDA will transform to LDA in cases where $\Tag$ of LDA is higher than that of HDA. Further warming will then melt LDA followed by rapid crystallization.  Indeed, a version of this predictable multi-step process has been observed experimentally\cite{amann2013water}.  The \emph{SI} illustrates this behavior with simulation trajectories made possible from our numerical preparation of HDA and LDA, and it further discusses this interpretation of the experiments.

\section{Preparations of amorphous ices with the s-ensemble.} 
 
As noted, amorphous ices (or any other glass) have structures distinct from those of an equilibrium liquid, distinct in the way excitations are distributed~\cite{keys2013calorimetric}. Preparation of amorphous ices in the laboratory can take microseconds to minutes to even hours.  This range of time scales required by experiment is inaccessible by straightforward molecular simulation.%
\footnote{Until this work, attempts at numerical simulations of HDA and LDA have ignored this time-scale issue, imagining that a glass is produced simply when a molecular dynamics trajectory seems non-ergodic over a relatively short simulation time.  For example, Giovambattista and co-workers\cite{chiu2013pressure} attempted to create HDA and LDA phases with cooling rates of $3\times10^{10}$\,K/s,  and they judged whether a glass is formed by examining changes in configurations over trajectories no longer than $10^{-7}$\,s.  In contrast, the required cooling rate to produce a reasonably stable glass, such as we  prepare and describe herein, is no faster than $10^8$\,K/s, and the time scales for aging such glass is no less than $10^{-4}$\,s.  See \emph{SI}.  Not surprisingly, the materials simulated in Ref.~\cite{chiu2013pressure} are not the HDA and LDA glasses prepared in the laboratory.  The alleged HDA phase of Ref.~\cite{chiu2013pressure}, for example, cannot transition to a lower-density material until decompressed to negative pressures, at which point it evaporates, while experimentally, HDA and LDA are robust and can interconvert reversibly around $p \approx 10^{3}$\,bar.} 
Nevertheless, it is possible to produce robust immobile amorphous states in a computer simulation.  It is done through an importance sampling that focuses on relevant parts of trajectory space.  The procedure is a nonequilibrium version of large-deviation formalism~\cite{touchette2009large}.  Such an approach has been successful in simulating stable glasses of simple-liquid mixtures~\cite{hedges2009dynamic,speck2012constrained, jack2011preparation,speck2012first}.  We adapt that approach here with one additional feature:  while employing a dynamical order parameter to highlight non-crystalline immobile states, as has been done before, we employ a second order parameter that distinguishes nonequilibrium immobile states of different densities.  Both order parameters are functions of path, as required to characterize nonequilibrium phases.  

The order parameter we use to measure mobility is the total number of enduring displacements (EDs) occurring in an $N$-particle system during a trajectory of length $\tobs$~\cite{speck2012constrained}.  Other functions of system history could also be used~\cite{hedges2009dynamic, jack2011preparation,speck2012first}.  An ED occurs when a particle jumps from one position to another, and it sticks for a significant period of time in the new position~\cite{elmatad2012manifestations}.  Such motions manifest the elementary excitations in a structural glass former~\cite{keys2011excitations}.  
They occur intermittently, and when one such event occurs, it takes on average $\Delta t$ to complete.  This instanton time, $\Delta t$, is much smaller than the structural relaxation time of a glass-forming melt.  Structural relaxation follows from coordinated motions of a large number of elementary excitations~\cite{keys2011excitations}.  

The number of EDs per particle per unit time is
\begin{equation}
\label{Eq:excite}
\hat{c}[\mathbf{x}(t)]  = \frac{\Delta t}{N\tobs} \sum_{i=1}^N \sum_{t=\Delta t}^{\tobs} \Theta\left(|\bar{\mathbf{r}}_i(t)-\bar{\mathbf{r}}_i(t - \Delta t ) | - a \right )\,,
\end{equation}
where $\mathbf{x}(t)$ stands for the trajectory of the system, $a$ is the displacement length (a fraction of a molecular diameter), $\Theta(x)$ is the unit Heaviside function,
and $\bar{\mathbf{r}}_i(t)$ is the position of molecule $i$, averaged over the time interval $t-\delta t/2$ to $t+\delta t/2$.  The averaging over $\delta t$ coarse-grains out non-enduring vibrations.  Applying the prescriptions of Ref.~\cite{keys2011excitations} to models of water gives $\Delta t$ as approximately the structural relaxation time at normal liquid conditions, and $\delta t$ an order of magnitude smaller.  For the calculations illustrated below, we use $\Delta t = 1$\,ps and $\delta t = 0.1$\,ps.  Other choices for $\Delta t$ and $\delta t$ yield consistent results.

The second order parameter we employ is a dimensionless measure of density history. For constant pressure and fixed $N$, it can be expressed in terms of the system's instantaneous density, $\rho(t)$:
\begin{equation}
\label{Eq:rho}
\hat{\rho}[\mathbf{x}(t)]  =  \frac{\Delta t}{\tobs} \sum_{t=\Delta t}^{\tobs} \frac{\rho(t-\Delta t) - \rho_\mathrm{xtl}}{\rho_\mathrm{liq}-\rho_\mathrm{xtl}} \, ,
\end{equation}
where $\rho_\mathrm{liq}$ and $\rho_\mathrm{xtl}$ are the average densities of the equilibrium liquid and crystal, respectively, at a particular thermodynamic state.

These order parameters have associated fields, which render the spatial patterns associated with distinct phases and interfaces.  The inter-excitation lengths, $\ell$ and $\ell_\mathrm{ne}$, characterize the patterns of the excitation field in the liquid and glass, respectively.

The relevant equilibrium probability distribution function is
\begin{equation}
\label{Eq:Peq}
P(c,\rho) = \langle \delta \left (c - \hat{c}[\mathbf{x}(t)] \right ) \delta \left (\rho - \hat{\rho}[\mathbf{x}(t)] \right ) \rangle_\mathrm{A} \, ,
\end{equation}
where $\delta(x)$ is Dirac's delta function and the subscripted angle brackets, $\langle \dots \rangle_\mathrm{A}$, denote equilibrium average over trajectories that include amorphous microstates only. Such microstates have small values of the Steinhardt-Nelson-Ronchetti $Q_6$ parameter~\cite{steinhardt1983bond}.  This parameter is finite for crystalline ice states and vanishes as $\mathcal{O}(1/\sqrt{N})$ for amorphous states.  It is therefore possible to identify reasonable ranges of $Q_6$ values that discriminate between amorphous and crystalline states of water.
The amorphous equilibrium distribution functional is $P[\mathrm{x}(t)] \propto p_\mathrm{eq}[\mathrm{x}(t)] \prod_t \Theta(Q_6^* - Q_6(x_t))$, where $p_\mathrm{eq}[\mathrm{x}(t)]$ is the unconstrained trajectory distribution, and $Q_6(x_t)$ is the crystalline order parameter for the system configuration at the $t-$th time interval.  We have checked that in the region of the equilibrium phase diagram where our calculations are performed that our results are insensitive to a cutoff, $Q_6^*$, to the extent that it is large enough to encompass typical liquid fluctuations and small enough to exclude crystal interface formation (i.e. for an $N=216$ particle system, the acceptable range is $0.1<Q_6^* <0.18$).  See Ref.~\cite{Limmer:2011p134503}.

Conditioned as it is to sample only amorphous states, $P(c,\rho)$ is unimodal, with the most probable region near the average values of $c$ and $\rho$ for the liquid. The distribution, however, exhibits fat tails at the low values of $c$ typical of glass.  These tails (i.e., large deviations) can be stabilized with nonequilibrium fields that couple to $\hat{c}[\mathbf{x}(t)]$ and $\hat{\rho}[\mathbf{x}(t)]$.  Specifically, with the fields $s$ and $\lambda$, the equilibrium distribution of trajectories, $P[\mathrm{x}(t)]$, is re-weighted to  
\begin{equation}
\label{Eq:reweight}
P_{s,\lambda}[\mathrm{x}(t)] \propto P[\mathrm{x}(t)]\,e^{-\{s  \hat{c}[\mathbf{x}(t)] - \lambda  \hat{\rho}[\mathbf{x}(t)]\} N \tobs},  
\end{equation}
for which the nonequilibrium order-parameter distribution is 
\begin{equation}
\label{Eq:PNeq}
P_{s,\lambda}(c,\rho)\propto P(c,\rho) \, e^{-(s c - \lambda \rho) N \tobs} \, . 
\end{equation}
Positive values of $s$ favor low-mobility (i.e., glassy) states, and positive values of $\lambda$ favor high-density states. 

We have applied these equations to the mW model of water~\cite{Molinero:2009p4008}.  The reference temperature and pressure of the mW model are $\To =250$\,K, and $\po=1$\,bar.  The mW model is the simplest of atomistic models to exhibit reversible thermodynamics, freezing and relaxation of water~\cite{Limmer:2011p134503, limmer2012phase,Molinero:2009p4008,moore2011structural,limmer2013corresponding}.  That it also faithfully exhibits transitions to and from glass, as we detail, is evidence that the model contains essential features underlying the physical properties of water both in and out of equilibrium.
\begin{figure*}
\begin{center}
\includegraphics[width=17cm]{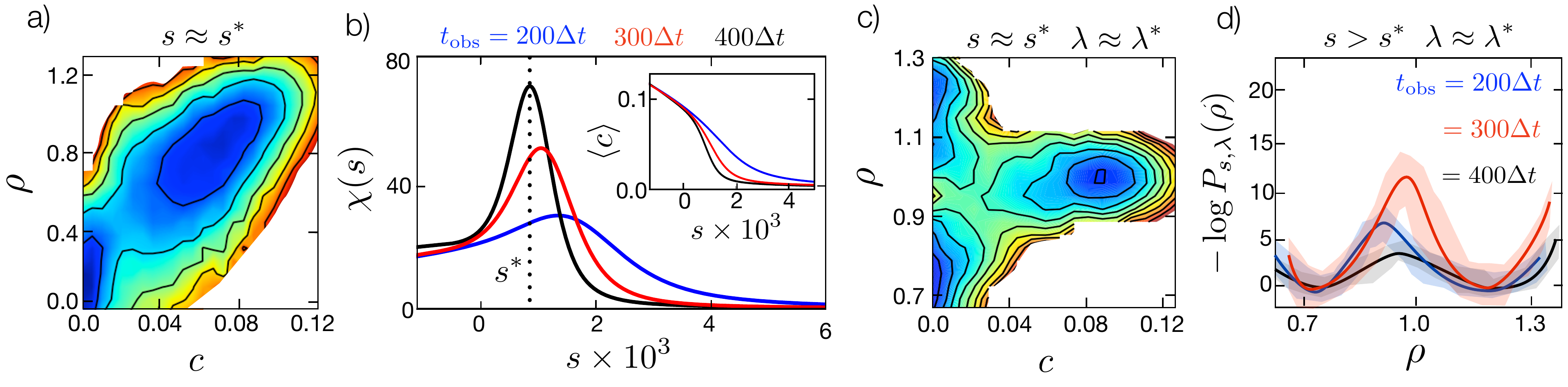}
\caption{\label{Fi:ldf}Nonequilibrium distributions for mobility, $c$, reduced density, $\rho$, and the susceptibility, $\chi(s)$, for cold water. (a) $ -\ln P_{s,\lambda}(c,\rho) $  calculated with the mW model for $t_\mathrm{obs}=300\Delta t$, $s\approx s^*$ and $\lambda=0$ at the state point $T/\To=0.8$, $p/p_\mathrm{o}=1$. 
(b) Mean mobility and susceptibility calculated at the state point in (a) for different trajectory lengths, $\tobs$, illustrating scaling consistent with a first-order phase transition in trajectory space.  The susceptibility peaks at nonequilibrium coexistence, $s=s^*$. (c) $-\ln P_{s,\lambda}(c,\rho)$ calculated with the mW model for $N=216$, $s\approx s^*$, $\lambda \approx \lambda^*$ and $t_\mathrm{obs}=200\Delta t$ at the state point $T/\To=0.75$, $p/p_\mathrm{o}=10^4$. (d) Marginal distribution functions of $\rho$ calculated for LDA-HDA coexistence at the state point in (c). Shading indicates error estimates of one standard deviation. Contours in (a) and (c) are spaced by unity, and the coloring is a guide to the eye.}
\end{center} 
\end{figure*}

Our trajectories fix the number of molecules, $N$, the pressure, $p$, and the temperature $T$. The system is evolved over a time $\Delta t$ with a Nose-Hoover barostat\cite{martyna1992nose}. At every $\Delta t$, all $N$-particle momenta are randomized, and this process is repeated up to a trajectory of length $\tobs$. We typically use $N=216$ and take $\tobs$ to be 10 to 400 times the structural relaxation time of the reversible melt.  The nonequilibrium distribution for these trajectories, Eq.~\ref{Eq:reweight}, is then sampled using transition path sampling~\cite{bolhuis2002transition}.  Reference~\cite{speck2012constrained} provides an illustration of such a calculation for a supercooled simple liquid mixture, but without the extra field $\lambda$.  The field $\lambda$ has a thermodynamic meaning, like a chemical potential, but affecting a time-averaged density rather than an instantaneous density.  In contrast, $s$ has a dynamical meaning, essentially the rate at which EDs are suppressed~\cite{garrahan2009first}.  

While this protocol overcomes huge time scales associated with glass transitions\cite{keys2014using}, the calculations are nevertheless time consuming.  As such, we have considered limited system sizes, large enough to exhibit clear signatures of glass transitions but not larger.  The side length of a simulation box with $N=216$ is slightly larger than $6 \sigma$, where $\sigma$ is a molecular diameter.  That side length is large compared to the equilibrium correlation length of the homogeneous liquid, which is about $\sigma$ or smaller.  But $6\sigma$ can be small compared to nonequilibrium lengths that characterize robust glasses.  Prior work~\cite{hedges2009dynamic, speck2012constrained} has found that anomalous responses of glass transitions begin to disappear from simulations when system sizes are decreased below 200 particles.  With $N\approx 200$, the stability of glasses we produce is limited to $\ell_\mathrm{ne} \approx 6 \sigma = 1.5$\,nm\cite{keys2014using}.

\section{Distinct phases and coexistence}
The nonequilibrium phase behavior we find in this way is illustrated in Figs.~\ref{Fi:noneqPD} and \ref{Fi:ldf}.  
We find three distinct amorphous phases: one ergodic liquid and two glasses.  For a finite $\tobs$ with fixed $p$ and $T$,  anomalous responses consistent with first-order transitions occur at specific values of $s$ and $\lambda$, which we label as $s^*$ and $\lambda^*$, respectively.  
Glasses formed at the higher temperatures require higher $s$ and are thus intrinsically less stable than those formed at lower $T$ with lower $s$.  The amorphous solid regions end where no value of $s$ can stabilize a glass distinct from the liquid.  That region cannot extend above $\To$.

The first-order characters of the glass transitions are manifested by precipitous changes in density and mobility that tend to discontinuities as $N \tobs \rightarrow \infty$.  Transitions between the two amorphous solids is illustrated in Fig.~\ref{Fi:noneqPD}b, and transitions between the amorphous solids and the liquid in Fig.~\ref{Fi:ldf}b. Consistent with experiments on salty water~\cite{bove2011pressure}, our coexistence line between the high density (HDA) and low density (LDA) solids ends at a triple point, not a critical point as supposed by Mishima~\cite{mishima1985apparently}. In a long trajectory at this nonequilibrium triple point, the system will visit each of the three phases and transition between them.  Figure~\ref{Fi:noneqPD}e shows a configuration near the triple point, transitioning between LDA and HDA.

From our explicit phase-coexistence calculations, like those illustrated in Fig.~\ref{Fi:ldf}, we have located the square points on Fig.~\ref{Fi:noneqPD}a.  These points lie in accord with the predictions of our analytical formulas for the glass transition temperature with $\ell_\mathrm{ne} = 6 \sigma = 1.5$\,nm.  This agreement provides numerical support for our understanding of the glass transition.  Further support comes from comparison with experiment.

The coexistence line between LDA and HDA occurs at the effective pressure $p-k_\mathrm{B}T\,\lambda^* \, \tobs /\Delta v = (5 \pm 3)\times 10^3 \, p_\mathrm{o}$.  (The uncertainty reflects the error estimates illustrated in Fig.~\ref{Fi:ldf}d.)  With $p_\mathrm{o} \approx 0.3$\,bar, the value of the reference pressure for water, the predicted coexistence is in harmony with experiments for the pressures found to produce reversible transitions between HDA and LDA~\cite{mishima1985apparently}.
The predicted density differences between LDA, HDA and liquid are also consistent with experiment within our corresponding states.  For example, converting the reduced density, $\rho$, to absolute experimental densities~\cite{eisenberg2005structure}, the results illustrated in Fig. 2 imply that at low pressures ($p/\po =1$) the density of the liquid is higher than that of LDA by 0.08 g/cm$^3$.  Similarly, at high pressures ($p/\po = 10^4$), the computed results imply that the density of HDA is higher than that of LDA by 0.12 g/cm$^3$; and at $p/\po =2\times10^4$, the computed results imply that the density of HDA is higher than that of the liquid by 0.005 g/cm$^3$.

The structure of the LDA glass is locally tetrahedral, as illustrated by the typical configuration shown in Fig.~\ref{Fi:noneqPD}d. The LDA basin has the same density as the crystalline phase, ordinary ice Ih, consistent with experimentally prepared LDA ices~\cite{debenedetti2003supercooled}. The local order is quantified with $\eta_{ijk} = (\mathbf{u}_{ij} \cdot \mathbf{u}_{ik} +1/3)^2$, where $\mathbf{u}_{ij}$ and  $\mathbf{u}_{ik}$ are the unit vectors pointing between a tagged molecule, $i$, to a pair of nearest neighbors, $j$ and $k$, respectively.
 For the LDA phase we have stabilized with the $s$-ensemble, $\langle \eta_{ijk} \rangle_\mathrm{A}\approx 0.05$.  In comparison, for the liquid and the HDA phase, $\langle \eta_{ijk} \rangle_\mathrm{A}\approx 0.2$.

HDA ice rendered in Fig.~\ref{Fi:noneqPD}f has an average structure similar to that of high pressure liquid water~\cite{soper2000structures}.  Our computed radial distribution functions for these phases are shown in Fig.~\ref{Fi:glass_relax}. The structures of the liquid and glass phases differ in the fluctuations from the average.  Spatial arrangements of excitations are uncorrelated in the liquid, but are anti-correlated with a large correlation length in a glass~\cite{keys2013calorimetric}.  This difference is most evident in the dynamics, Fig.~\ref{Fi:noneqPD}c, because the anti correlation arrests mobility~\cite{keys2013calorimetric,sollich1999glassy}.  Notice from the plateau values of $F_s(k,t)$ that fluctuations in molecular positions in HDA are larger than those in LDA.  This juxtaposition predicted from our simulations is consistent with experiment\cite{amann2012limits}.

The marginal distribution of $c$, $\int \mathrm{d} \rho \, P_{s, \lambda}(c,\rho) $, has mean value, 
$\langle c \rangle$, and its variance gives the susceptibility, $\chi(s) =  -(\partial \langle c \rangle/\partial s)_\lambda =N \tobs \langle (c -\langle c \rangle)^2 \rangle $.  In the thermodynamic limit, $\langle c \rangle$ and $\chi(s)$
are singular functions at the point of a glass transition, $s=s^*$.  In simulations, the development of this singular behavior can be detected from system-size dependence. Specifically, for a first-order transition, the width of the change in $\langle c \rangle$ around $s=s^*$ should decrease proportionally to $1/N\tobs$, and the height of $\chi(s)$ at $s=s^*$ should grow proportionally to $N\tobs$.  This scaling with respect to space-time volume is exhibited by the functions graphed in Fig.~\ref{Fi:ldf}b.  Similarly, at coexistence, the free energy barrier between the two stable basins should grow proportionally to space-time surface area, $(N\tobs)^{3/4}$.  This scaling is consistent with the growth exhibited in Fig.~\ref{Fi:ldf}d, though a compelling demonstration is beyond the scope of the small system size and statistics we are able to treat. Also, as space and time obey different symmetries, finite size scaling may depend on other combinations of $N$ and $t_\mathrm{obs}$. See, for example, analogous issues in theory of quantum phase transitions\cite{cardy1996scaling}.

\section*{Melting and transitioning between amorphous solids}
Having prepared glassy configurations with the $s$-ensemble, we can now study two experimental observations. 
The first is the non-monotonic thermal responses found when heating LDA. The material first takes in heat, then it precipitously releases heat and crystallizes~\cite{elsaesser2010reversibility, loerting2006amorphous}.  The experimental LDA coincides with the LDA that is first prepared with the $s$-ensemble at some temperature $T<\To$ and then cooled to a lower temperature where it remains stable for essentially all time.  Melting LDA occurs when that low temperature is increased, a process that can be simulated by simply turning off $s$ at the initial preparation temperature.

\begin{figure}[h]
\begin{center}
\includegraphics[width=8.0cm]{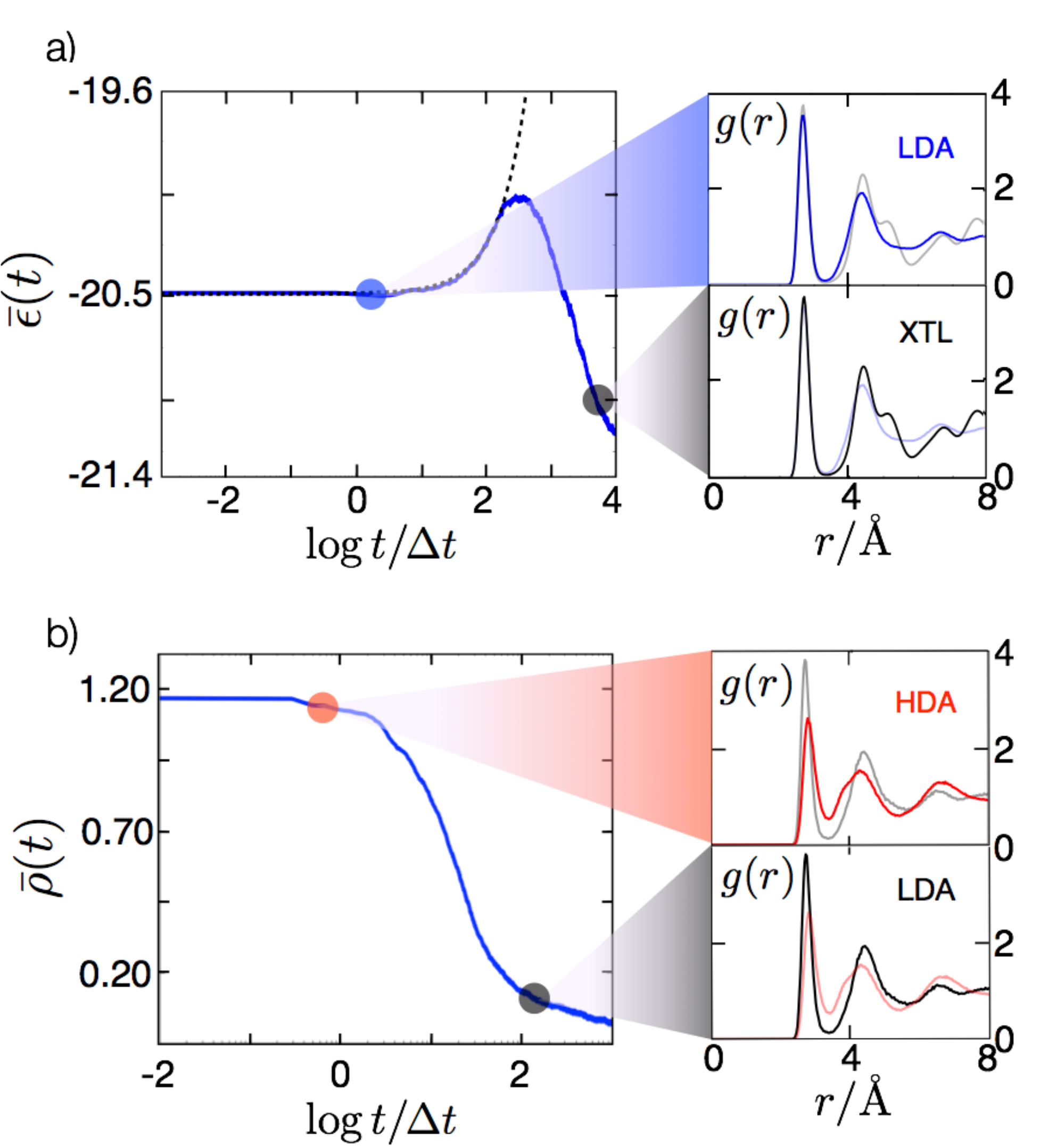}
\caption{\label{Fi:glass_relax}Relaxation behavior of amorphous ices produced with the $s$-ensemble. (a) Average potential energy per particle, in units of $\To$, as a function of time for the mW model prepared in an ensemble at $s>s^*$, $T/\To=0.8$, $p/p_\mathrm{o}=0$ and $t_\mathrm{obs}=200\Delta t$ and evolved with $s=0$, $T/\To=0.76$, $p/p_\mathrm{o}=0$. The dashed black line is an exponential function with characteristic time, $200\Delta t$. Right panels shown the average pair distribution functions at two indicated points in time. Faint lines show the $g(r)$ for the alternative solid.  (b) Average reduced density as a function of time for the mW model prepared in an ensemble at $s>s^*$, $T/\To=0.76$, $p/p_\mathrm{o}=2 \times10^4$ and $t_\mathrm{obs}=200\Delta t$ and evolved with $s=0$, $T/\To=0.6$, $p/p_\mathrm{o}=5\times 10^3$. Right panels show the average pair distribution functions at the two indicated points in time.  Faint lines show the $g(r)$ for the alternative solid.}
\end{center} 
\end{figure}

Results of such simulations are shown in Fig.~\ref{Fi:glass_relax}.  The nonequilibrium average potential energy per molecule in units of $T_\mathrm{o}$, $\bar{\epsilon}(t)$,  is computed by averaging over 1000 independent trajectories initiated from configurations taken from the ensemble of inactive states.  With $s=0$, these amorphous solid states are thermodynamically unstable.  The stable basin is the crystal, but that basin cannot be accessed without reorganization, and reorganization requires access to ergodic liquid states.  The inactive glassy states are at a low potential energy state relative to the supercooled liquid. Upon instantaneously turning off the $s$-field, the system remains immobile for a relatively long time, on average about $t=200\Delta t$.  This waiting time corresponds to the time for a rare fluctuation to produce an excitation. Once this reorganization begins, the system immediately begins to crystallize, and by $t=1000\Delta t$ on average the system has begun releasing energy as long-ranged order builds up. The right panels of Fig.~\ref{Fi:glass_relax}a show the average radial distribution functions, $g(r)$, for the beginning and end of the trajectory. Initially, the radial distribution function shows the local order characteristic of LDA, indicated by the separation between the first and second solvation shell~\cite{finney2002structures}. At the end of the trajectory, this local ordering has developed into a long ranged ordered crystal, as indicated by the splitting of the second solvation shell and the persistent correlations at large separations.

The second experimental observation we consider is the finding of an abrupt transition from HDA to LDA when HDA is quenched to lower pressures keeping temperature low~\cite{mishima1985apparently}.  This process can be simulated by initiating trajectories at configurations from an immobile HDA basin, prepared with $s>s^*$ and $p/p_\mathrm{o}>10^4$, and running these trajectories with $s=0$ and $p/p_\mathrm{o}<10^4$. Figure \ref{Fi:glass_relax}b shows the result from averaging over 1000 such trajectories.  The average waiting time to transition across the HDA-LDA boundary is only $10 \Delta t$, reflecting that only relatively small reorganization is required for transitioning between these two amorphous phases. 
The excess free energy due to the change in pressure is dissipated through an average concentration of mobility, $c$, that is only 0.02. After the initial burst of excitation, the system monotonically relaxes into the low density amorphous state. 
Initially, the structure reflects the HDA configurations where the dynamics were initialized, while at later times the structure adopts the open local order of LDA.  

Other illustrations of behaviors deduced from our preparations of amorphous ices are given in the \emph{SI}.  For example, reversal and hysteresis of the process illustrated in  Fig.~\ref{Fi:glass_relax}b is shown, demonstrating that the glassy states prepared in our simulations are robust.  No prior simulations of low temperature water have achieved this quality. 

\section{Conclusions}
The most important and general results of this work are two-fold: the demonstration that it is possible with molecular simulation to systematically prepare and predict properties and transitions of experimentally realizable amorphous solids, and the demonstration of analytical theory that can predict and interpret various behaviors of these materials. We have thus illustrated new possibilities for molecular simulation and theory.  For water in particular, we present here the first prediction and simulation of LDA-HDA transitions at conditions consistent with experimental observations.  We also present the first prediction of density differences between LDA, HDA and ergodic liquid phases in accord with experimental observations; and finally, we present predictions of pathways by which HDA and LDA phases melt, again in accord with experimental observations.    

Much has been written suggesting that the HDA-LDA transition might reflect a transition between two distinct phases of liquid water, e.g.,  Refs.~\cite{angell2004amorphous, mishima1985apparently,Debenedetti:2003p813, angell1995formation,mishima1998relationship,poole1994effect}.  This extrapolation from glass to liquid seems difficult to justify in light of the singularity that separates the nonequilibrium amorphous solids from the ergodic liquid.  Occurring as they do through driving a material out of equilibrium over a finite period of time, the space-time transition is precipitous but not discontinuous, but a singularity underlies the phenomena nonetheless.  This fact about glass physics may have its first experimental demonstration in Ref.~\cite{bove2011pressure}, where coexistence lines for reversible transitions between different glasses of salty water are established and shown to not extend above the line of glass transition temperatures.
 
That particular experimental work finds more than one coexistence line separating distinct amorphous phases.  Our discussion of phenomenology emphasizes that any one material can exhibit a range of glass behaviors reflecting a range of values of $\ell_\mathrm{ne}$.  How this variability can translate in general into distinct nonequilibrium phases is not yet known, and the answer is likely system dependent.  For example, distinct amorphous phases seem generally possible in cases of a poor glass forming liquid, as it is for water, because coarsening of crystal phases of one density can compete with vitrification of the liquid of another density.  Whether other competing effects can be imagined and realized experimentally is an open question.

\begin{acknowledgments} 
We thank Juan P. Garrahan, Robert L. Jack, Osamu Mishima and Thomas Speck for comments on early drafts.  The Helios Solar Energy Research Center, which was supported by the Director, Office of Science, Office of Basic Energy Sciences of the U.S. Department of Energy under Contract No. DE-AC02-05CH11231 provided salaries;  NSF award CHE-1048789 provided computational resources.
\end{acknowledgments}


\end{article}

\clearpage





\renewcommand{\thefigure}{S\arabic{figure}}

\begin{article}

\noindent
{\huge\bf Supporting Information} 
\\
{\Large\bf for ``Theory of Amorphous Ices''} 

\bigskip
\noindent
{\large\bf by D. T. Limmer and D. Chandler }

\setcounter{figure}{0} \renewcommand{\thefigure}{S\arabic{figure}}
\setcounter{equation}{0} \renewcommand{\theequation}{S\arabic{equation}}

\section{Glass transition temperatures and nonequilibrium lengths}

This section of the \emph{SI} summarizes formulas we use in this paper to predict glass transition temperatures.  The formulas are derived in Refs.~\cite{keys2013calorimetric1} and \cite{limmer2013communication1}.

Heterogeneous dynamics below the onset is characterized by the concentration, $c$, of localized soft spots or excitations.  At equilibrium, $c\sigma^3=\exp(-\tilde{\beta})$, where $\tilde{\beta} = [1/T - 1/T_\mathrm{o}(p)]J_\sigma(p) > 0$ and $J_\sigma(p)$ is the free energy or reversible work to move a molecule a molecular diameter $\sigma$.  The pressure dependence of $J_\sigma(p)$ and $\To(p)$ are important when considering the behavior of water.  

While dynamics above the onset temperature is unstructured, like random motion in a mean field, dynamics below the onset temperature is controlled by excitations facilitating the birth and death of neighboring excitations.  At equilibrium, the mean-free path between excitations is
\begin{equation}
\ell(T) = \sigma \exp(\tbeta/\df)\,,
\label{Eq:ell}
\end{equation}
where $\df$ is the fractal dimensionality of the path.  For three-dimensional structural glass, $\df \approx 2.4$\cite{keys2011excitations1}.

Collective reorganization is required to move a molecule to a new enduring position, so that its reorganization energy depends upon the length of that displacement.  Specifically, $J_{\sigma'} = J_\sigma \,[1 + \gamma \ln(\sigma'/\sigma)]$.  This logarithmic growth of energy with length is universal, but the constants $J_\sigma$ and $\gamma$ are system dependent~\cite{Elmatad:2009p75331,keys2011excitations1}.  As a result of the logarithmic growth, the structural relaxation time, $\tau$, is
\begin{equation}
\tau(T) = \tau_\mathrm{MF}\, \exp \{\tbeta \,\gamma \ln [\ell(T)/\sigma] \} \,,\quad T<\To(p), 
\label{Eq:tau}
\end{equation}
where $\tau_\mathrm{MF}$ is $\tau$ for $T\geqslant \To(p)$.  In general, $\tau_\mathrm{MF}$ is a weak function of $T$ and $p$, but we neglect that dependence in comparison with the much larger temperature variation of the right-hand side of Eq.~\ref{Eq:tau}.  At equilibrium, Eqs.~\ref{Eq:ell} and \ref{Eq:tau} combine to give the familiar super-Arrhenius parabolic law.

Super-Arrhenius relaxation is associated with underlying hierarchical dynamics, where relaxation depends upon the size of relaxing domains.  This dependence is responsible for a glass transition when the material is cooled at a rate $\nu$.  Specifically, the system transitions from ergodic to non-ergodic behavior at a temperature $\Tg$, where 
\begin{equation}
1/\nu = |d\tau/dT|_{\Tg}\,,
\label{Eq:GlassT}
\end{equation} 
below which $\ell(T)$ is locked at its nonequilibrium value $\ell_\mathrm{ne} = \ell(\Tg)$.  
Therefore
\begin{equation}
1 = \frac{2\, \nu \,\tau_\mathrm{MF}\,\tilde{\beta}_\mathrm{g}\gamma \,J_\sigma}{\df \, \Tg^2}\,\exp\left( \tilde{\beta}_\mathrm{g}^2 \gamma /\df \right)\,,
\label{Eq:trans}
\end{equation}
where 
\begin{equation}
\tilde{\beta}_\mathrm{g} =J_\sigma(p) [1/\Tg(p) - 1/\To(p)] =  \df\, \ln(\ell_\mathrm{ne}/\sigma)\,,
\end{equation}
or
\begin{equation}
\frac{1}{\Tg(p)} = \frac{1}{\To(p)} + \frac{\df}{J_\sigma(p)} \ln(\ell_\mathrm{ne}/\sigma)\,.
\label{Eq:TgEll}
\end{equation}
Equation \ref{Eq:TgEll} gives the dashed lines in Fig. 1a.
  
An approximate solution to the transcendental  Eq.~\ref{Eq:trans} is useful when $\To$ and $\Tg$ are of the same order and $ \tilde{\beta}_\mathrm{g}^2 \gamma /\df \gg \ln( \tilde{\beta}_\mathrm{g} \gamma /\df)$.  In that case 
\begin{equation}
\ln(\ell_\mathrm{ne}/\sigma) \approx \sqrt{- \ln \left [2 \, \nu \,\tau_\mathrm{MF} \,\gamma\,J_\sigma (p)/\To^2 (p) \right ] / \df \gamma\,}\,.
\end{equation} 
This solution can serve as the first guess to the numerical solution, the first guess differing from the numerical solution for water by a few percent.

Because $\ell(T) = \ell_\mathrm{ne}$ for all $T<\Tg$, the relaxation time $\tau$, Eq.~\ref{Eq:tau}, is Arrhenius for that regime.  If the glass with its fixed $\ell_\mathrm{ne}$ is cooled to a very low temperature, and then warmed on a time scale of $t_\mathrm{w}$, it will undergo a transition at an apparent glass transition temperature $\Tag$, where  $t_\mathrm{w} \, = \,\tau_\mathrm{MF}\, \exp \{\tbeta_\mathrm{ag} \,\gamma \ln (\ell_\mathrm{ne}/\sigma) \}$.  Accordingly,
\begin{equation}
\frac{1}{\Tag(p)} = \frac{1}{\To(p)} + \frac{\ln(t_\mathrm{w} / \tau_\mathrm{MF})}{\gamma \, J_\sigma(p)  \, \ln(\ell_\mathrm{ne}/\sigma)} \,  .
\label{Eq:Tag}
\end{equation}
Equation \ref{Eq:Tag} yields the dotted lines in Fig. 1a, with $\Tag(p)$ evaluated for a warming time scale of minutes, i.e., $t_\mathrm{w} \approx 10^{2}\, \mathrm{s} \approx 10^{14}\,\tau_\mathrm{MF}$.

\begin{figure}
\begin{center}
\includegraphics[width=8.5cm]{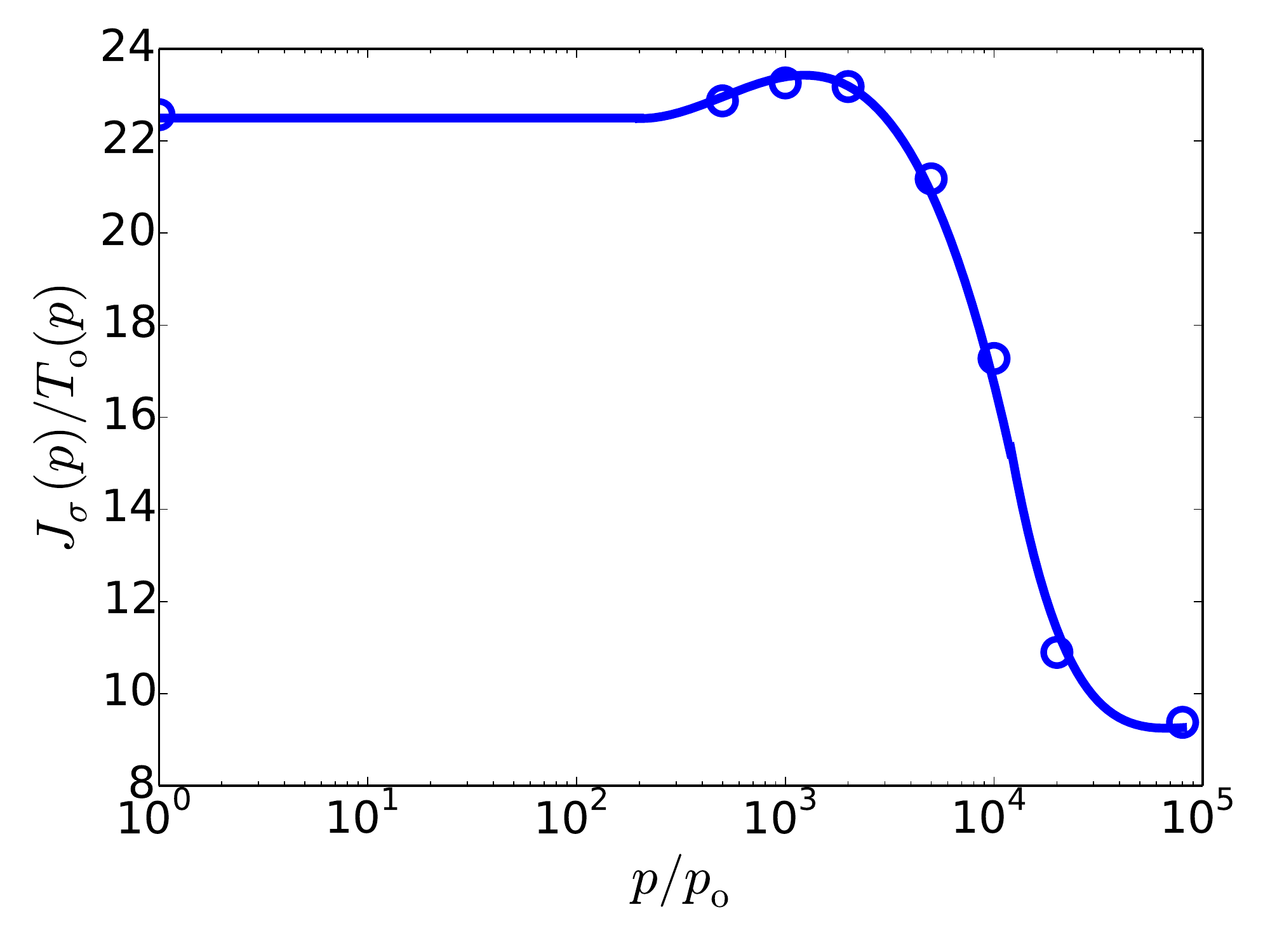}
\caption{Reduced energy scale, $J_\sigma(p)/\To(p)$, as a function of pressure.  Circles are results computed with the mW model following methods detailed in Refs.~\cite{keys2011excitations1} and \cite{limmer2012phase1}.  Corresponding states analysis~\cite{limmer2013corresponding1} indicates that these results should hold for all reasonable models of water as well as for the actual substance.  The solid line is the spline fit to the data.
\label{Fi:JofP}}
\end{center} 
\end{figure}

Application of these formulas require $\To(p)$, $J_\sigma(p)$  and $\tau_\mathrm{MF}$. The low pressure forms have been determined previously\cite{limmer2012phase1}. High pressure behaviors have been determined similarly. Figure 1a shows the behavior of $\To(p)$, and its form is well approximated by a spline,
\begin{eqnarray}
\To(p)/\To &=& -0.015 \log(p/\po) \nonumber \\
&& +0.976 \, , \, 0<p/\po<5 \times10^2 \nonumber \\
&=& 0.199 \log^2(p/\po) \nonumber \\
&& -1.344 \log(p/\po) \,,\nonumber \\
&&  + 3.118 ,   \, \, 5\times10^2<p/\po<2\times10^3\nonumber \\
&=& 0.173(\log^{2}(p/\po) \nonumber \\
&& -1.078 \log(p/\po) \nonumber \\
&& +2.521\, , \, \, 2\times10^3<p/\po<8\times10^4\nonumber \, .  \\
\end{eqnarray}
Figure \ref{Fi:JofP} shows the behavior of $J_\mathrm{\sigma}(p)/\To(p)$, and its form is well approximated by a spline,
\begin{eqnarray}
J_\mathrm{\sigma}(p)/\To(p) &=& 22.5  \, , \, 0<p/\po<2 \times10^2 \nonumber \\
&=& -3.9(\log(p/\po)-2.7)^3 \nonumber \\
&& +1.8 \log(p/\po) \,,\nonumber \\
&&  +18.1,   \, \, 2\times10^2<p/\po<1.2\times10^4\nonumber \\
&=& 6.0(\log(p/\po)-5.1)^4 \nonumber \\
&& +0.5 \log(p/\po) \nonumber \\
&& +6.8 \, , \, \, 1.2\times10^4<p/\po<8\times10^4\nonumber \, .  \\
\end{eqnarray}

Table S1 illustrates predictions of these formulas applied to water at ambient conditions, computing $\Tag$ with the warming-time scale of minutes, i.e., $t_\mathrm{w} = 10^2$\,s. 

\begin{table}[h]
\fignumfont{Table~S1. }\figtextfont{Nonequilibrium length, time and energy scales for LDA ice at ambient pressure.}
\tabletextfont
\begin{center}
\begin{tabular}{ l c c c r}
$ \ell_\mathrm{ne}/$nm & $\nu/\mathrm{K \,s}^{-1}$  & $\tau_\mathrm{g}/\mathrm{s}$  &  $T_\mathrm{g}/\To$ & $T_\mathrm{ag}/\To$  \\
 \hline 
1.5  & $10^8$ & $10^{-9}$ & 0.80 & 0.48 \\
5.0 & $10^4$ & $10^{-4}$ & 0.73 & 0.55 \\
10.0  & 0.1 & $10^{2}$ & 0.65 & 0.65  \\ 
\end{tabular}
\end{center}
\begin{minipage}{0.4\columnwidth}
\end{minipage}
\end{table}

\section*{Dynamics of transformations of amorphous ices}

In this section of the \emph{SI} we provide a few more examples of dynamics that follow from our simulated HDA and LDA phases.

\begin{figure}[h]
\begin{center}
\includegraphics[width=8.5cm]{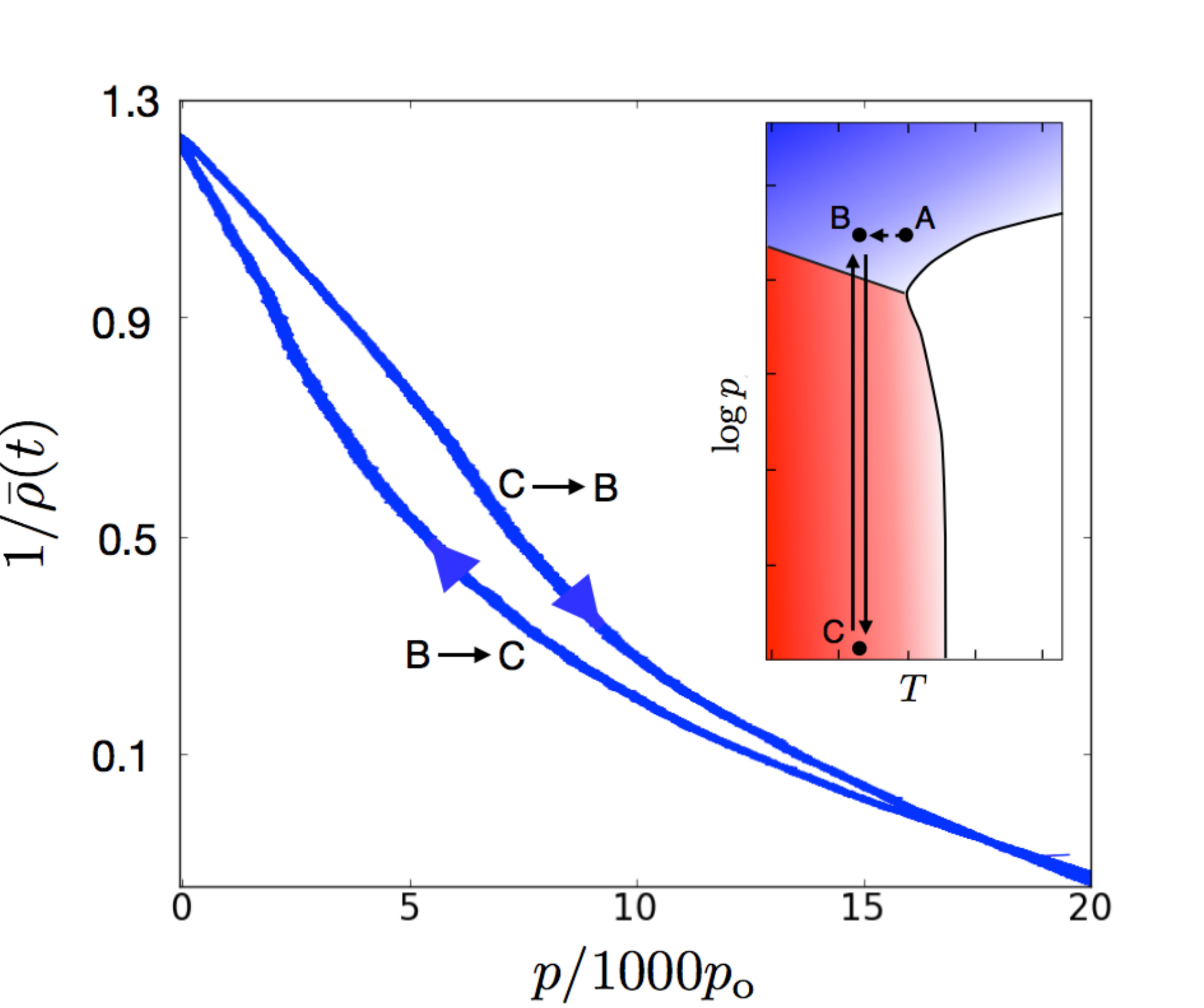}
\caption{
Forward and backward transitions between HDA and LDA. Configurations taken from HDA prepared at $s>s^{*}$, $T/\To=0.8$ and $p/\po=20\times 10^3$, state A, are instantly quenched at constant pressure to a temperature $T/\To=0.6$, state B, where it is annealed with $s=0$ for 2000 $\Delta t$. Then the pressure is changed at constant temperature at a rate of -5 kbar/ns to $p/\po$=1, state C. The pressure is then changed at constant temperature with a rate of 5 kbar/ns, back to state B. The paths are illustrated in the inset, and the time dependence of the averaged reduced density or volume is illustrated in the main graph.
\label{Fi:LDAHDA_Reverse}}
\end{center} 
\end{figure}

The first example focuses on the reversibility of pressurizing and depressurizing the amorphous ices to transition between HDA and LDA.  The nature of these processes is illustrated with Fig.~\ref{Fi:LDAHDA_Reverse}.  Specifically, configurations taken from the HDA basin prepared with large $s$ are first quenched to lower temperature and to $s=0$. Then the configuration, and volume are evolved with Nose-Hoover\cite{martyna1992nose1} equations of motion with a constant rate of change of the pressure and its reverse. Over 1000 trajectories generated in this way are used to compute the time dependent density depicted in Fig.~\ref{Fi:LDAHDA_Reverse}. The ability to reverse the HDA to LDA transition demonstrates that the materials produced by the $s$-ensemble are robust solids.  

The second example considers the time-dependence of the potential energy per particle, $\bar{\epsilon}(t)$, and the number of EDs per particle, $\bar{c}(t)$, of very cold HDA brought to a low pressure where it is then warmed.  From our discussion of phenomenology in the main text and from experimental work~\cite{amann2012limits1}, we expect this protocol to produce two calorimetric peaks -- one apparent glass transition where HDA transforms to LDA, and another apparent glass transition where LDA melts into a non-equilibrium liquid from which crystal ice coarsens.

\begin{figure}[h]
\begin{center}
\includegraphics[width=8.5cm]{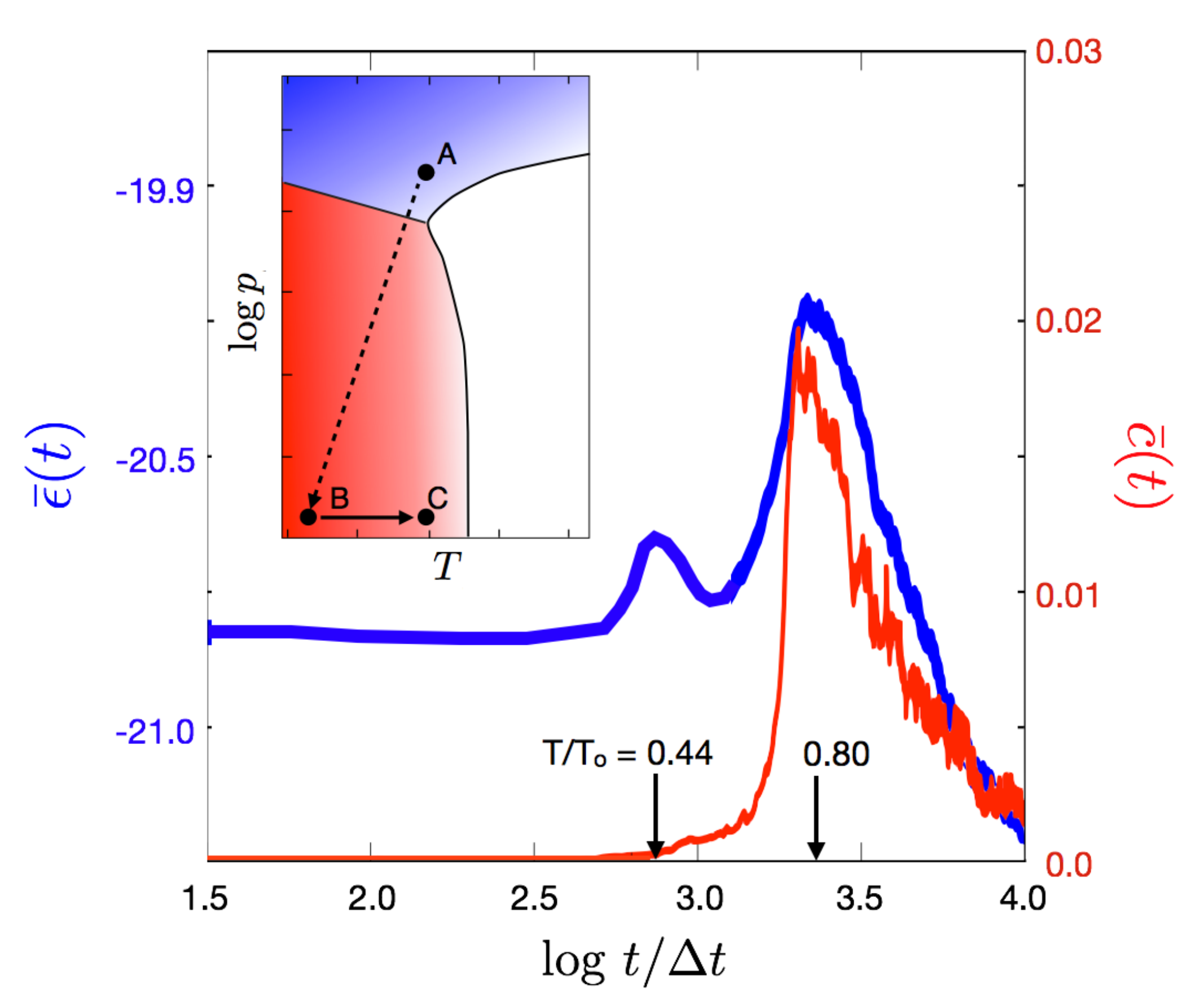}
\caption{Time dependence of the potential energy and number of enduring displacements of HDA heated at low pressure. Configurations taken from HDA prepared at $s>s^{*}$, $T/\To=0.8$ and $p/\po=20\times 10^3$, state A, are instantly quenched to $s=0$, $T/\To=0.32$ and $p/\po=1$, state B. The temperature is then changed at constant pressure at a rate of 10 K/ns to $T/\To=0.8$, state C. Configurations are then annealed at this temperature for $t/\Delta t= 7.5 \times 10^3$. This path is illustrated in the inset.\label{Fi:Loerting}, and the time dependence of the averaged potential energy per particle, $\bar{\epsilon}(t)$ in units of $\To$, and the excitation concentration, $\bar{c}(t)$ are shown in the main graph.  The black arrows indicate the temperature reached at two particular points in time.}
\end{center} 
\end{figure}

We observe this behavior, as illustrated in Fig. \ref{Fi:Loerting}, and the temperatures at which the transitions occur can be understood in terms of the equations presented in the previous section.  The figure shows the results obtained from averaging 1,000 independent trajectories initiated from the HDA configurations, with a warming-time scale $t_\mathrm{w} \approx 10^3 \tau_\mathrm{MF}$.  Equation \ref{Eq:Tag} then predicts a transition at $\Tag \approx 0.44\, \To$, in good agreement where the low-temperature transition is detected in the trajectories.  Above that temperature, the radial distribution functions found from our simulation indicate that the resulting amorphous solid is the LDA material.  In that case, the activation energy (or equivalently, the value of $J_\sigma$) has changed from that locked in from the higher pressure HDA material to that of the LDA material.  Equation~\ref{Eq:Tag} then gives $\Tag \approx 0.80$ for the temperature that LDA will melt, again in good agreement with the results of our trajectories.

Having gained confidence in our theoretical analysis through comparison with simulation, we now turn to the experimental observations of two-step relaxation~\cite{amann2013water1}.  These recent experiments have found that a stabilized version of HDA brought to low temperatures and pressures exhibits a calorimetric peak at $T \approx 130$\,K.  By taking this value for $\Tag$ and applying Eq.~\ref{Eq:Tag} with $t_\mathrm{w} = 100$\,s, we conclude that $\ell_\mathrm{ne} \approx 5$\,nm for this version of HDA.  The corresponding $\Tag$ for LDA can then be predicted using this value for $\ell_\mathrm{ne}$ together with the low pressure LDA value for $J_\sigma$.  This evaluation predicts $\Tag \approx 150$\,K for the LDA material that is produced by melting the stabilized HDA.  This predicted position for a second calorimetric peak is in harmony with experiment~\cite{amann2013water1}.

Notice that had experimentalists not stabilized the HDA through annealing, the data shown in Fig. 1a suggests that the HDA material would have $\ell_\mathrm{ne} \approx 1.5$\,nm.  In that case, $\Tag$ computed from Eq.~\ref{Eq:Tag} for that HDA material would be $\approx 85$\,K.  Such a low value for the temperature at which the low pressure form of HDA would become unstable indicates why two-step melting was not detected without first annealing to create a more stable HDA.  

Based upon indirect evidence, experimentalists have interpreted two-step melting of amorphous ices as indicative of two distinct liquid phases~\cite{amann2012limits1, amann2013water1}.  We find nothing in our simulations to support the idea. The time dependence of the excitation concentration, $\bar{c}(t)$, shows that the material remains solid like until reaching the apparent glass transition temperature of LDA, which with the warming rate of our simulations occurs near $0.8\,\To$.  In other words, some reorganization does occur to allow the transition from HDA to LDA, but the low mobility of a glass remains until ergodic states are accessed at the apparent glass transition temperature for LDA.  


\end{article}

\end{document}